\documentclass[article,12pt]{revtex4}
\usepackage{amsfonts}
\usepackage{amssymb}
\usepackage{txfonts}
\usepackage{appendix}
\usepackage{calligra}
\usepackage{times}

\DeclareMathAlphabet{\mathcalligra}{T1}{calligra}{m}{n}
\DeclareFontShape{T1}{calligra}{m}{n}{<->s*[2.2]callig15}{}
\newcommand{\scripty}[1]{\ensuremath{\mathcalligra{#1}}\;}

\begin{document}

\title{Geometric phases in astigmatic optical modes of arbitrary order}

\author{Steven J.M. Habraken}
\email{habraken@molphys.leidenuniv.nl}
\author{Gerard Nienhuis}
\address{Leiden Institute of Physics, P.O. Box 9504, 2300 RA Leiden, The Netherlands}
\begin{abstract}
The transverse spatial structure of a paraxial beam of light is fully characterized by a set of parameters that vary only slowly under free propagation. They specify bosonic ladder operators that connect modes of different order, in analogy to the ladder operators connecting harmonic-oscillator wave functions. The parameter spaces underlying sets of higher-order modes are isomorphic to the parameter space of the ladder operators. We study the geometry of this space and the geometric phase that arises from it. This phase constitutes the ultimate generalization of the Gouy phase in paraxial wave optics. It reduces to the ordinary Gouy phase and the geometric phase of non-astigmatic optical modes with orbital angular momentum states in limiting cases. We briefly discuss the well-known analogy between geometric phases and the Aharonov-Bohm effect, which provides some complementary insights in the geometric nature and origin of the generalized Gouy phase shift. Our method also applies to the quantum-mechanical description of wave packets. It allows for obtaining complete sets of normalized solutions of the Schr\"odinger equation. Cyclic transformations of such wave packets give rise to a phase shift, which has a geometric interpretation in terms of the other degrees of freedom involved.
\end{abstract}

\maketitle

\section{Introduction}
In the twenty-five years that have passed since Berry published his landmark paper \cite{Berry84}, the geometric phase has turned out to be a very unifying concept in physics. Various phase shifts and rotation angles both in classical and quantum physics have been proven to originate from the geometry of the underlying parameter space. One of the first examples was given by Pancharatnam \cite{Pancharatnam56} who discovered that the phase shift due to a cyclic transformation of the polarization of an optical field is equal to half the enclosed area on the Poincar\'e sphere for polarization states. Other optical examples of geometric phases are the phase shift that arises from the variation of the direction of the wave vector of an optical field through a fiber \cite{Tomita86} and the phase that is associated with the cyclic manipulation of a squeezed state of light \cite{Chiao88}. The Gouy phase shift, which is due to the variation of the beam parameters (the beam width and the radius of curvature of the wave front) of a Gaussian optical beam, can also be interpreted geometrically \cite{Simon93}.

In analogy with the geometric phase for polarization (or spin) states of light, van Enk has proposed a geometric phase that arises from cyclic mode transformations of paraxial optical beams carrying orbital angular momentum \cite{Enk93}. The special case of first-order modes is equivalent to the polarization case \cite{Padgett99} and, as was experimentally demonstrated by Galvez et. al., the geometric phase shift acquired by a first-order mode that is transformed along a closed trajectory on the corresponding Poincar\'e sphere also equals half the enclosed surface on this sphere \cite{Galvez03}. Similar experiments have been performed with second-order modes \cite{Galvez05}, in particular to show that exchange of orbital angular momentum is necessary for a non-trivial geometric phase to occur. However, in the general case of isotropic modes of order $N$, the connection with the geometry of the $N+1$-dimensional mode space is not at all obvious. Moreover, although astigmatic lenses are essential for mode conversion \cite{Beijersbergen93}, the geometric interpretation in \cite{Enk93} and \cite{Galvez05} applies only to set-ups where the initial and final modes are non-astigmatic.

In this paper, we present a complete and general analysis of the phase shift of, possibly astigmatic, transverse optical modes of arbitrary order when propagating through a paraxial optical set-up, thereby resolving the above-mentioned issues. Paraxial optical modes with different transverse mode indices $(n,m)$ are connected by bosonic ladder operators in the spirit of the algebraic description of the quantum-mechanical harmonic oscillator and complete sets of transverse modes $|u_{nm}\rangle$ can thus be obtained from two pairs of ladder operators \cite{Visser04}. We show that the geometries of the subspaces of modes with fixed transverse mode numbers $n$ and $m$, which are closed under mode transformations, are all isomorphic to the geometry underlying the ladder operators. We fully characterize this geometry including both the generalized beam parameters, which characterize the astigmatism and orientation of the intensity and phase patterns of a Gaussian fundamental mode, and the degrees of freedom associated with the nature and orientation of the higher-order modes. We find a dynamical and a geometric contribution to the phase shift of a mode under propagation through an optical set-up, which both have a clear interpretation in terms of this parameter space.

The material in this paper is organized as follows. In the next section we briefly summarize the operator description of paraxial wave optics. We discuss its group-theoretical structure, which is essential for our ladder-operator approach, and show how paraxial ray optics emerges from it. In section \ref{6Basis sets of paraxial modes} we discuss how complete basis sets of transverse modes can be obtained from two pairs of bosonic ladder operators. We discuss the transformation properties of the ladder operators, and, thereby, of the modes and we characterize the ten degrees of freedom that are associated with the choice of a basis of transverse modes. Two of those degrees of freedom relate to overall phase factors of the ladder operators and, therefore, of the modes. In section \ref{6The geometric origin of variation of the phases}, we show that the variation of these phases under propagation through a set-up originates from the variation of the other parameters. The well-known analogy between geometric phases and the Aharonov-Bohm effect provides an alternative way to derive an expression for the variation of the phases. This complementary approach, which is also discussed in section \ref{6The geometric origin of variation of the phases}, shows unambiguously that both contributions to the phase shift are geometric in that they are fully determined by the trajectory through the parameter space. Moreover, it provides a deeper geometric understanding of the dependence  of the phase shifts of the modes on the transverse mode numbers $n$ and $m$. In the final section, we summarize our results, draw our conclusions and discuss the relevance of our present work in the context of quantum mechanics.

\section{Canonical description of paraxial optics}
\label{6Canonical description of paraxial optics}
\subsection{Position and propagation direction as conjugate variables}
A monochromatic paraxial beam of light that propagates along the $z$ direction is conveniently described by the complex scalar profile $u(\rho,z)$, which characterizes the spatial structure of the field beyond the structure of the carrier wave $\exp(ikz-i\omega t)$. The two-dimensional vector $\rho=(x,y)^{\mathrm{T}}$ denotes the transverse coordinates. The electric and magnetic fields of the beam can be expressed as
\begin{equation}
\label{6efield}
\mathbf{E}(\rho,z,t)=\mathrm{Re}\left\{E_{0}\epsilon u(\rho,z)e^{ikz-i\omega t}\right\}
\end{equation}
and
\begin{equation}
\label{6bfield}
\mathbf{B}(\rho,z,t)=\mathrm{Re}\left\{\frac{E_{0}}{c}(\mathbf{e}_{z}\times\epsilon) u(\rho,z)e^{ikz-i\omega t}\right\}\;,
\end{equation}
where $E_{0}$ is the amplitude of the field, $\epsilon$ is the transverse polarization, $\mathbf{e}_{z}$ is the unit vector along the propagation direction and $\omega=ck$ is the optical frequency
with $c$ the speed of light. The slowly varying amplitude $u(\rho,z)$ obeys the paraxial wave equation
\begin{equation}
\label{6pwe}
\left(\nabla_{\rho}^{2}+2ik\frac{\partial}{\partial z}\right)u(\rho,z)=0\;,
\end{equation}
where $\nabla_{\rho}^2=\partial^2/\partial x^2+\partial^2/\partial y^2$ is the transverse Laplacian. Under the assumption that the transverse variation of the field appears on a much larger length
scale than the wavelength, this description of paraxial wave optics is consistent with Maxwell's equations in free space \cite{Lax75}.

The paraxial wave equation (\ref{6pwe}) has the form the Schr\"odinger equation for a free particle in two dimensions. The longitudinal coordinate $z$ plays the role of time while the transverse coordinates $\rho=(x,y)^{\mathrm{T}}$ constitute the two-dimensional space. This analogy allows us to adopt the Dirac notation of quantum mechanics to describe the evolution of a classical wave field \cite{Stoler81}. In the Schr\"odinger picture, we introduce state vectors $|u(z)\rangle$ in the Hilbert space $L^{2}$ of square-integrable transverse states of the wave field, where the $z$ coordinate parameterizes the trajectory along which the field propagates. The states are properly normalized $\langle u(z)|u(z)\rangle=1$ for all $z$ and the field profile in real space can be expressed as $u(\rho,z)=\langle\rho|u(z)\rangle$. Just as in quantum mechanics, the transverse coordinates may be viewed as a hermitian vector operator $\hat{\rho}=(\hat{x},\hat{y})^{\mathrm{T}}$ acting on the Hilbert space. The derivatives with respect to these coordinates constitute canonically conjugate operators. Rather than the conjugate transverse momentum operator $-i\partial/\partial \rho$, which has the significance of the normalized transverse momentum of the field, it is convenient to construct the propagation-direction operator by dividing the transverse momentum operator by the normalized longitudinal momentum $k$. Thus, we obtain the hermitian vector operator $\hat{\theta}=(\hat{\vartheta}_{x},\hat{\vartheta}_{y})^{\mathrm{T}}=-(i/k)(\partial/\partial x,\partial/\partial y)^{\mathrm{T}}$. The transverse position and propagation-direction operators obey the canonical commutation rules
\begin{equation}
\label{6cancom}
[\hat{\rho}_{a},k\hat{\theta}_{b}]=i\delta_{ab}\;,
\end{equation}
where the indices $a$ and $b$ run over the $x$ and $y$ components. In analogy with quantum mechanics, we introduce the transverse field profile in propagation-direction representation
\begin{equation}
\tilde{u}(\theta,z)=\langle\theta|u(z)\rangle= \frac{k}{2 \pi}\int d_2\rho\;u(\rho,z) e^{-ik\theta^{\mathrm{T}}\rho}\;,
\end{equation}
which is the two-dimensional Fourier transform of $u(\rho,z)$ and characterizes the transverse propagation-direction distribution of the field.

In geometric optics, a ray of light is fully characterized in a transverse plane $z$ by its transverse position $\rho$ and propagation direction $\theta$, which are usually combined in the four-dimensional ray vector $\scripty{r}^{\mathrm{T}}=\big(\rho^{\mathrm{T}},\theta^{\mathrm{T}}\big)$. The operator description of paraxial wave optics may be viewed as a formally quantized (wavized) description of light rays, where $\rho$ and $\theta$ have been replaced by hermitian operators $\hat{\rho}$ and $\hat{\theta}$ that obey canonical commutation rules (\ref{6cancom}) and $1/k=\lambdabar$ plays the role of $\hbar$ \cite{Gloge69}. These operators are conveniently combined in the ray operator $\hat{\scripty{r}}^{\mathrm{T}}=\big(\hat{\rho}^{\mathrm{T}},\hat{\theta}^{\mathrm{T}}\big)$. In analogy with quantum mechanics, where the expectation values of the position and momentum operators have a clear classical significance in the limit $\hbar\rightarrow 0$, a paraxial wave field reduces to a ray in the limit of geometric optics $\lambdabar\rightarrow 0$. Its transverse position and propagation direction in the transverse plane $z$ are characterized by the expectation values $\langle u(z)|\hat{\rho}|u(z)\rangle$ and $\langle u(z)|\hat{\theta}|u(z)\rangle$.

\subsection{Group-theoretical structure of paraxial wave and ray optics}
Both the diffraction of a paraxial beam under free propagation, as described by the paraxial wave equation (\ref{6pwe}), and the transformations due to lossless optical elements can be expressed as unitary transformations $|u_{\mathrm{out}}\rangle=\hat{U}|u_{\mathrm{in}}\rangle$ on the transverse state of the field. In general, a unitary operator can be expressed as
\begin{equation}
\label{6U}
\hat{U}\big(\{a_{j}\}\big)=e^{-i\sum_{j}a_{j}\hat{T}_{j}}\;,
\end{equation}
where $\{a_{j}\}$ is a set of real parameters and $\{\hat{T}_{j}\}$ a set of hermitian generators, i.e., $\hat{T}_{j}^{\dag}=T_{j}$. In the present case of paraxial propagation and paraxial (first-order) optical elements, the generators are quadratic forms in the transverse position and propagation-direction operators. This is exemplified by the paraxial wave equation (\ref{6pwe}), which in operator notation takes the following form
\begin{equation}
\frac{\partial}{\partial z} |u(z)\rangle = - \frac{ik}{2}\hat{\theta}^2 |u(z)\rangle
\end{equation}
and is formally solved by
\begin{equation}
\label{6formsolpwe}
|u(z)\rangle=\exp\left(-\frac{ikz\hat{\theta}^2}{2}\right)|u(0)\rangle\;.
\end{equation}
This shows that that free propagation of a paraxial field is generated by $k\hat{\theta}^2/2$, which is obviously quadratic in the canonical operators. Since the ray operator $\hat{\scripty{r}}$ has four components, the number of squares of the operators is four while the number of mixed products is ${4 \choose 2} = 6$, which gives a total of ten quadratic forms. They are hermitian and can be chosen as
\begin{eqnarray}
\label{6generators}
&T_{1}=\hat{x}^2\;,\quad T_{2}=\hat{y}^2\;,\quad T_{3}=\hat{x}\hat{y}\;,\nonumber\\
&T_{4}=\frac{k}{2}\left(\hat{x}\hat{\vartheta}_{x}+\hat{\vartheta}_{x}\hat{x}\right)\;,\quad T_{5}=\frac{k}{2}\left(\hat{y}\hat{\vartheta}_{y}+\hat{\vartheta}_{y}\hat{y}\right)\;, \quad T_{6}=k\hat{x}\hat{\vartheta}_{y}\;,\quad T_{7}=k\hat{y}\hat{\vartheta}_{x}\;,\nonumber\\
&T_{8}=k^{2}\hat{\vartheta}_{x}\hat{\vartheta}_{y}\;,\quad T_{9}=k^{2}\hat{\vartheta}^{2}_{x}\;\quad\mathrm{and}\quad T_{10}=k^{2}\hat{\vartheta}^{2}_{y}\;.
\end{eqnarray}
In terms of these generators, free propagation of a paraxial beam (\ref{6formsolpwe}) is described by
\begin{equation}
\label{6freeprop}
|u(z)\rangle=\exp\left(-\frac{i\left(\hat{T}_{9}+\hat{T}_{10}\right)z}{2k}\right)|u(0)\rangle\;.
\end{equation}
The mixed product $\hat{T}_{8}$ appears in the generator of free propagation through an anisotropic medium, i.e., a medium in which the refractive index depends on the propagation direction $\theta$. In that case the propagator can be expressed as $\exp(-ik\hat{\theta}^{\mathrm{T}}\mathsf{N}^{-1}\hat{\theta}z/2)$, where $\mathsf{N}$ is a real and symmetric matrix that characterizes the (quadratic) variation of the refractive index with the propagation direction. If the anisotropy of the refractive index is not aligned along the $\vartheta_{x}$ and $\vartheta_{y}$ directions, this transformation also involves $\hat{T}_{8}$. A thin astigmatic lens imposes a Gaussian phase profile. The unitary transformation that describes it can be expressed as
\begin{equation}
\label{6astigmaticlens}
|u_{\mathrm{out}}\rangle=\exp\left(-\frac{ik\rho^{\mathrm{T}}\mathsf{F}^{-1}\rho}{2}\right)|u_{\mathrm{in}}\rangle\;,
\end{equation}
where $\mathsf{F}$ is a real and symmetric $2\times 2$ matrix whose eigenvalues correspond to the focal lengths of the lens while the corresponding, mutually perpendicular, eigenvectors fix its orientation in the transverse plane. In the general case of an astigmatic lens that is not aligned along the $x$ and $y$ directions, this transformation involves the generators $\hat{T}_{1}$, $\hat{T}_{2}$ and $\hat{T}_{3}$. A rotation of the beam profile in the transverse plane can be represented by
\begin{equation}
\label{6rotation}
|u_{\mathrm{rot}}\rangle=e^{-i(\hat{T_{6}}-\hat{T}_{7})\phi}|u\rangle\;,
\end{equation}
where $\hat{T_{6}}-\hat{T}_{7}=-i(x\partial/\partial y-y\partial/\partial x)$ is the orbital angular momentum operator and $\phi$ is the rotation angle. The operators $\hat{T_{4}}$ and $\hat{T}_{5}$ generate transformations that rescale a field profile along the $x$ and $y$ directions respectively, i.e.,
\begin{equation}
\label{6rescale}
u_{\mathrm{out}}(x,y,z)=\langle\rho|u_{\mathrm{out}}(z)\rangle=\langle\rho|e^{i\log(c_{x})\hat{T}_{4}+i\log(c_{y})\hat{T}_{5}}|u_{\mathrm{in}}(z)\rangle=\sqrt{c_{x}c_{y}}\;u_{\mathrm{in}}(c_{x}x,c_{y}y,z)\;.
\end{equation}
Physically speaking, such transformations correspond to the deformation of a field profile due to refraction at the interface between two dielectrics with different refractive indices.

From the canonical commutation relations (\ref{6cancom}), it follows that the commutator of any two generators (\ref{6generators}) is a linear combination of the generators. In mathematical terms, the algebra of the generators is closed, which means that $[\hat{T}_{k},\hat{T}_{l}]=i\sum_{m}g_{klm}\hat{T}_{m}$ with real structure constants $g_{klm}$. We shall prove that the unitary transformations (\ref{6U}) with the generators (\ref{6generators}) form a ten-parameter Lie group. For reasons that will become clear this group is called the metaplectic group $Mp(4)$.

Since the states $|u(z)\rangle$ are normalized, the expectation values $\langle u(z)|\hat{\rho}|u(z)\rangle$ and $\langle u(z)|\hat{\theta}|u(z)\rangle$ have the significance of the average transverse position and the average propagation direction of the field. A special property of the unitary transformations in equation (\ref{6U}) with the quadratic generators given by (\ref{6generators}), is that the Heisenberg transformation $\hat{U}^{\dag}\hat{\scripty{r}}\hat{U}$ of the vector operator $\hat{\scripty{r}}^{\mathrm{T}}= \big(\hat{\rho}^{\mathrm{T}},\hat{\theta}^{\mathrm{T}}\big)$ is linear, so that it can be expressed as
\begin{equation}
\label{6ehrenfest}
\hat{U}^{\dag}\big(\{a_{j}\}\big)\hat{\scripty{r}}\hat{U}\big(\{a_{j}\}\big)=M\big(\{a_{j}\}\big)\hat{\scripty{r}}\;,
\end{equation}
where $M\big(\{a_{j}\}\big)$ is the $4\times 4$ ray matrix that describes the transformation of a ray $\scripty{r}^{\mathrm{T}}=\big(\rho^{\mathrm{T}},\theta^{\mathrm{T}}\big)$ under the optical element that is described by the state-space operator $\hat{U}\big(\{a_{j}\}\big)$. The defining properties of the position and momentum operators, i.e., that they are hermitian and obey canonical commutation rules (\ref{6cancom}), are preserved under this unitary Heisenberg transformation. It follows that $M\big(\{\alpha_{j}\}\big)$ is real and obeys the identity
\begin{equation}
\label{6mgm}
M^{\mathrm{T}}\big(\{a_{j}\}\big)GM\big(\{a_{j}\}\big)=G\qquad\mathrm{with}\qquad G=\left(\begin{array}{cc}\mathsf{0}&\mathsf{1}\\-\mathsf{1}&\mathsf{0}\end{array}\right)\;,
\end{equation}
where $\mathsf{0}$ and $\mathsf{1}$ denote the $2\times 2$ zero and unit matrices respectively, so that $G$ is a $4 \times 4$ matrix. This identity (\ref{6ehrenfest}) ensures that the operator expectation values $\langle u(z)|\hat{\scripty{r}}|u(z)\rangle$ of the transverse position and propagation direction transform as a ray, i.e., trace out the path of a ray when the field propagates through an optical set-up. This shows how paraxial ray optics emerges from paraxial wave optics and, as such, the identity (\ref{6ehrenfest}) may be viewed as an optical analogue of the Ehrenfest theorem in quantum mechanics. The manifold of rays $\scripty{r}$ constitutes a phase space in the mathematical sense. The real and linear transformations on this manifold that obey the relation (\ref{6mgm}), or, equivalently, preserve the canonical commutation rules (\ref{6cancom}), are ray matrices. The product of two ray matrices is again a ray matrix so that ray matrices form a group. The group of real $4\times 4$ ray matrices, which preserve the bilinear form $\scripty{r}^{\mathrm{T}}G\scripty{s}$, where $\scripty{r}$ and $\scripty{s}$ are ray vectors, is called the symplectic group $Sp(4,\mathbb{R})$. The $4\times 4$ ray matrices in $Sp(4,\mathbb{R})$ emerge from a set of unitary state-space transformations, which, as one may prove from equation (\ref{6ehrenfest}), constitute a group under operator multiplication. As was mentioned already, this group is called the metaplectic group $Mp(4)$. For real rays $\scripty{r},\scripty{s}\in\mathbb{R}^{4}$, the products $\scripty{r}^{\mathrm{T}}G\scripty{r}$ and $\scripty{s}^{\mathrm{T}}G\scripty{s}$ vanish. The product $\scripty{r}^{\mathrm{T}}G\scripty{s}$ does not vanish and is obviously conserved under paraxial propagation and optical elements. It is called the Lagrange invariant and has the significance of the phase-space extent of a pair of rays $\scripty{r}$ and $\scripty{s}$. Conservation of this quantity is an optical analogue of the Liouville theorem in statistical mechanics.

The commutators of the quadratic generators $\hat{T}_{j}$ and the position and propagation-direction operators are linear in these operators, so that we can write
\begin{equation}
\label{6homomorphism}
-i[\hat{T}_{j},\hat{\scripty{r}}]=J_{j} \hat{\scripty{r}}\;,
\end{equation}
where the $4 \times 4$ matrices $J_j$ are real. Explicit expressions of these matrices are given in appendix \ref{6The ray-space generators}. Applying equation (\ref{6ehrenfest}) to infinitesimal transformations immediately shows that the ray matrix corresponding to the unitary state-space operator in equation (\ref{6U}) is given by
\begin{equation}
\label{6M}
M\big(\{\alpha_{j}\}\big)=e^{-\sum_{j}\alpha_{j}J_{j}}\;.
\end{equation}
Equation (\ref{6homomorphism}) provides a general relationship between the generators $\{\hat{T}_{j}\}$ of the unitary state-space transformations (\ref{6U}) and the generators $\{J_{j}\}$ of the corresponding ray matrices (\ref{6M}). By applying equation (\ref{6mgm}) to infinitesimal transformations, one finds that the generators obey $J^{\mathrm{T}}_{j}G+GJ_{j}=0$. Moreover, from equation (\ref{6homomorphism}) one may prove that
\begin{equation}
\big[[\hat{T}_{i},\hat{T}_{j}], \hat{\scripty{r}}\big]=[J_{i},J_{j}]\hat{\scripty{r}}\;.
\end{equation}
Using the Lie algebra $[\hat{T}_{k},\hat{T}_{l}]=i\sum_{m}g_{klm}\hat{T}_{m}$ we find that $[J_{k},J_{l}]=-\sum_{m}g_{klm}J_{m}$. This proves that the metaplectic and symplectic groups are homomorphic, i.e., for every $\hat{U}\in Mp(4)$ there is a corresponding $M\in Sp(4,\mathbb{R})$. The reverse of this statement is not true; a ray matrix $M$ fixes a corresponding transformation $\hat{U}$ up to an overall phase. The homomorphism is an isomorphism up to this phase.

By using equation (\ref{6ehrenfest}) and the expressions of the unitary transformations (\ref{6freeprop}), (\ref{6astigmaticlens}), (\ref{6rotation}) and (\ref{6rescale}) or, equivalently, the relation (\ref{6homomorphism}) between the sets of generators $\{\hat{T}_{j}\}$ and $\{J_{j}\}$ and the definition of the ray matrices (\ref{6M}), one finds the $4\times 4$ ray matrices that describe propagation, a thin lens, a rotation in the transverse plane and the rescaling of a beam profile due to refraction at the interface between two dielectrics. These ray matrices, some of which are given explicitly in \cite{Habraken07} and \cite{Habraken08}, generalize the well-known ABCD matrices to the case of two independent transverse degrees of freedom \cite{Siegman}.

The group-theoretical structure that we have discussed in this section can easily be generalized to the case of $D$ spatial dimensions. In that case there are $2D$ canonical operators. These give rise to $2D+{2D\choose 2}= 2D^2+D$ linearly independent quadratic forms, which generate state-space transformations that constitute the metaplectic group $Mp(2D)$. The corresponding ray matrices obey the $2D-$dimensional generalization of equation (\ref{6mgm}) and form the corresponding symplectic group $Sp(2D,\mathbb{R})$. In case of a single transverse dimension, the three hermitian quadratic forms can be chosen as $x^2$, $k(\hat{x}\hat{\vartheta}_{x} +\hat{\vartheta}_{x}\hat{x})/2$ and $k^2\hat{\vartheta}_{x}^{2}$. In the analogous case of the quantum-mechanical description of a particle in three dimensions, the number of quadratic forms is twenty-one.

\section{Basis sets of paraxial modes}
\label{6Basis sets of paraxial modes}
\subsection{Ladder operators}
As a result of the quadratic nature of the generators (\ref{6generators}), a, possibly astigmatic, Gaussian beam profile at the $z=0$ input plane of a paraxial optical set-up will retain its Gaussian shape in all other transverse planes $z$. This is the general structure of a transverse fundamental mode. Complete sets of higher-order transverse modes that preserve their general shape under paraxial propagation and paraxial optical elements can be obtained by repeated application of bosonic raising operators $\hat{a}_p^{\dag}(0)$ in the $z=0$ plane \cite{Nienhuis93}. In the present case of two transverse dimensions, we need two independent raising operators so that $p=1,2$. Both the raising operators and the corresponding lowering operators $\hat{a}_p(0)$ are linear in the transverse position and propagation-direction operators $\hat{\rho}$ and $\hat{\theta}$. Their transformation property under unitary transformations $\in Mp(4)$ follows from the requirement that acting with a transformed ladder operator on a transformed state must be equivalent to transforming the raised or lowered state, i.e.,
\begin{equation}
\hat{a}^{(\dag)}_{\mathrm{out}}|u_{\mathrm{out}}\rangle=\hat{a}^{(\dag)}_{\mathrm{out}}\hat{U}|u_{\mathrm{in}}\rangle=\hat{U}\hat{a}^{(\dag)}_{\mathrm{in}}|u_{\mathrm{in}}\rangle\;.
\end{equation}
In view of the unitarity of $\hat{U}$, this requires that
\begin{equation}
\label{6ladoptrans}
\hat{a}^{(\dag)}_{\mathrm{out}}=\hat{U}\hat{a}^{(\dag)}_{\mathrm{in}}\hat{U}^{\dag}\;.
\end{equation}
Since the generators (\ref{6generators}) are quadratic in the position and propagation-direction operators, the ladder operators preserve their general structure and remain linear in these operators under this transformation (\ref{6ladoptrans}). Moreover, their bosonic nature is preserved so that they obey the commutation rules
\begin{equation}
\label{6boscom}
[\hat{a}_{p}(z),\hat{a}^{\dag}_{q}(z)]=\delta_{pq}
\end{equation}
in all transverse planes $z$ of the optical set-up if (and only if) they obey bosonic commutation rules in the $z=0$ plane. When the fundamental Gaussian mode $|u_{00}(z)\rangle$ is chosen such that the lowering operators give zero when acting upon it, i.e., $\hat{a}_{1}(z)|u_{00}(z)\rangle = \hat{a}_{2}(z)|u_{00}(z)\rangle=0$, the commutation rules (\ref{6boscom}) guarantee that the modes
\begin{equation}
\label{6modes}
|u_{nm}(z)\rangle=\frac{1}{\sqrt{n!m!}}\left(\hat{a}_{1}^{\dag}(z)\right)^{n}\left(\hat{a}_{2}^{\dag}(z)\right)^{m}|u_{00}(z)\rangle\;,
\end{equation}
form a complete set in all transverse planes $z$. For a given optical system, the complete set of modes is thus fully characterized by the choice of the two bosonic ladder operators $\hat{a}_p(0)$ in the reference plane $z=0$.

In reference \cite{Habraken07}, we have shown that, in the special case of an astigmatic two mirror-cavity, the ladder operators, and thereby the cavity modes, can be directly obtained as the eigenvectors of the ray matrix for one round trip inside the cavity. In the present case of an open system, we are free to choose the parameters that specify the ladder operators in the $z=0$ input plane. A convenient way to do this is to choose an arbitrary ray matrix $M_{0}\in Sp(4,\mathbb{R})$. This ray matrix can be chosen independent of the properties of the optical system, and of the ray matrices that describe the transformations of its elements. However, as we shall see, a necessary and sufficient restriction is that $M_{0}$ has four eigenvectors $\mu$ for which the matrix element $\mu^{\dag}G\mu$ does not vanish. It is obvious that this matrix element is purely imaginary so that the eigenvectors must be complex. Since $M_{0}$ is real, this implies that for each eigenvector $\mu_{p}$ also $\mu_{p}^{\ast}$ is one of the eigenvectors so that the eigenvectors come in two complex-conjugate pairs, obeying the eigenvalue relations $M_{0}\mu_p = \lambda_p \mu_p$ and $M_{0} \mu_p^{\ast}= \lambda_p^{\ast}\mu_p^*$, with $p=1,2$. Without loss of generality we can assume that the matrix elements $\mu_p^{\dag}G \mu_p$ are positive imaginary. Then we can write
\begin{equation}
\label{6muprop}
\mu_{p}^{\dag}G\mu_{p}=2i\qquad\mathrm{and}\qquad\mu_{p}^{\mathrm{T}}G\mu_{p}=0\;,
\end{equation}
where $p=1,2$. The first relation can be assured by proper normalization of the eigenvectors, whereas the second follows from the antisymmetry of $G$. By taking matrix elements of the symplectic
identity $M_{0}^\mathrm{T}GM_{0} = G$, we find the relations
\begin{equation}
\label{6mugmu}
\lambda_p^{\ast} \lambda_q\mu_{p}^{\dag}G\mu_{q}=\mu_{p}^{\dag}G\mu_{q}\qquad\mathrm{and}\qquad\lambda_p \lambda_q\mu_{p}^{\mathrm{T}}G\mu_{q}=\mu_{p}^{\mathrm{T}}G\mu_{q}\;.
\end{equation}
Assuming that the two eigenvalues $\lambda_1$ and $\lambda_2$ are different, we conclude that
\begin{equation}
\label{6mugmu2}
\mu_{1}^{\dag}G\mu_{2}=0\qquad\mathrm{and}\qquad\mu_{1}^{\mathrm{T}}G\mu_{2}=0\;.
\end{equation}
When the eigenvalues are degenerate, i.e., $\lambda_{1}=\lambda_{2}$, one can find infinitely many pairs of linearly independent vectors $\mu_{1}$ and $\mu_{2}$ that obey these symplectic orthonormality properties. Following the approach discussed in \cite{Habraken07}, we now specify the ladder operators in the $z=0$ input plane by the expressions
\begin{equation}
\label{6ladderoperators}
\hat{a}_p(0)=\sqrt{\frac{k}{2}}\mu_p^{\mathrm{T}}G\hat{\scripty{r}}\qquad\mathrm{and}\qquad\hat{a}_p^{\dag}(0)=\sqrt{\frac{k}{2}}\mu_p^{\dag}G\hat{\scripty{r}}\;.
\end{equation}
The symplectic orthonormality properties (\ref{6muprop}) and (\ref{6mugmu2}) of the eigenvectors $\mu_{p}$ and $\mu_{p}^{\ast}$ ensure that the ladder operators in the input plane obey bosonic commutation relations (\ref{6boscom}). From the general transformation property of the ladder operators (\ref{6ladoptrans}), combined with the Ehrenfest relation (\ref{6ehrenfest}) between $\hat{U}$ and $M$, one may show that the ladder operators in other transverse planes $z$ are given by the same expressions (\ref{6ladderoperators}) when $\mu_{p}$ is replaced by $\mu_{p}(z)=M(z)\mu_{p}$. Here, $M(z)$ is the ray matrix that describes the transformation of a ray from the $z=0$ input plane to the transverse plane $z$. It can be constructed by multiplying the ray matrices that describe the optical elements of which the set-up consists and free propagation between them in proper order. The fact that the properties (\ref{6muprop}) and (\ref{6mugmu2}) are conserved under symplectic transformations $\in Sp(4,\mathbb{R})$ confirms that the ladder operators remain bosonic in all transverse planes of the set-up.

Since the modes are fully characterized by the choice of two complex vectors $\mu_{p}$, we expect that the expectation values of physically relevant operators can be expressed in terms of these vectors. The average transverse position and momentum of the beam trace out the path of a ray. This implies that the expectation values $\langle u_{nm}|\hat{\rho}|u_{nm}\rangle$ and $\langle u_{nm}|\hat{\theta}|u_{nm}\rangle$ vanish. In appendix \ref{6Expectation values of the generators} we prove, however, that the expectation values of the generators $\hat{T}_{j}$ are, in general, different from zero and can be expressed as
\begin{equation}
\label{6unmTJ}
\langle u_{nm}|\hat{T}_{j}|u_{nm}\rangle=\frac{1}{2}\left\{\left(n+\frac{1}{2}\right)\mu_{1}^{\dag}GJ_{j}\mu_{1}+\left(m+\frac{1}{2}\right)\mu_{2}^{\dag}GJ_{j}\mu_{2}\right\}\;.
\end{equation}

Finally, it is worthwhile to notice that the results of this section remain valid when the number of (transverse) dimensions is different. In particular, the same method gives explicit expressions for complete orthogonal sets of time-dependent wave functions that solve the Schr\"odinger equation of a free particle in three-dimensional space.

\subsection{Degrees of freedom in fixing a set of modes}
We have shown that there is a one-to-one correspondence between the defining properties of a ray matrix, i.e., that it is real and obeys the identity (\ref{6mgm}), and the properties (\ref{6muprop}) and (\ref{6mugmu2}) of the complex eigenvectors $\mu_{p}$ that ensure that the ladder operators (\ref{6ladderoperators}) are bosonic. This implies that all different basis sets of complex vectors $\mu_{p}$ that obey these identities must be related by symplectic transformations, i.e., each of these sets can be written as $\{M\mu_{p}\}\cup \{M\mu_{p}^{\ast}\}$, with $M\in Sp(4,\mathbb{R})$ and $\{\mu_{p}\}\cup\{\mu_{p}^{\ast}\}$ the set of complex eigenvectors of a specific ray matrix $M_{0}\in Sp(4,\mathbb{R})$. Since $\{M\mu_{p}\}\cup \{M\mu_{p}^{\ast}\}$ constitutes the set of eigenvectors of $MM_{0}M^{-1}$, it follows that the freedom in choosing a set of complex vectors that generate two pairs of bosonic ladder operators (\ref{6ladderoperators}) is equivalent to the freedom of choosing a ray matrix $M\in Sp(4,\mathbb{R})$. As a result, the number of independent parameters associated with this choice is equal to the number of generators of $Sp(4,\mathbb{R})$, which is ten. In order to give a physical interpretation of these degrees of freedom, we decompose the complex ray vectors into two-dimensional subvectors so that $\mu_{p}^{\mathrm{T}}(z)=\big(r_{p}^{\mathrm{T}}(z),t_{p}^{\mathrm{T}}(z)\big)$. In terms of these subvectors, the ladder operators (\ref{6ladderoperators}) take the following form
\begin{equation}
\label{6ladderoperators2}
\hat{a}_p(z)=\sqrt{\frac{k}{2}}\big(r_p^{\mathrm{T}}(z)\hat{\theta} - t_p^{\mathrm{T}}(z)\hat{\rho}\big)\qquad\mathrm{and}\qquad\hat{a}_p^{\dag}(z)=\sqrt{\frac{k}{2}}\big(r_p^{\dag}(z)\hat{\theta} - t_p^{\dag}(z) \hat{\rho}\big)\;,
\end{equation}
where $p=1,2$. An explicit expression of the Gaussian fundamental mode can be given if we combine the two-dimensional column vectors $r_p$ and $t_p$ into
\begin{equation}
\label{6RTdef}
\mathsf{R}(z)=\big(r_1(z),r_2(z)\big)\qquad\mathrm{and}\qquad\mathsf{T}(z)=\big(t_1(z),t_2(z)\big)\;.
\end{equation}
The objects $\mathsf{R}$ and $\mathsf{T}$ take the form of $2\times 2$ matrices, but since $r_{p}$ and $t_{p}$ are transverse vectors, $\mathsf{R}$ and $\mathsf{T}$ do not transform as such under ray-space transformations $\in Sp(4,\mathbb{R})$ nor under transformations on the transverse plane. The symplectic orthonormality properties (\ref{6muprop}) and (\ref{6mugmu2}) of the vectors $\mu_{p}$ can be expressed as
\begin{equation}
\label{6RTprop}
\mathsf{R}^{\dag}(z) \mathsf{T}(z)-\mathsf{T}^{\dag}(z)\mathsf{R}(z)=2i\mathsf{1}\qquad\mathrm{and}\qquad\mathsf{R}^{\mathrm{T}}(z)\mathsf{T}(z)-\mathsf{T}^{\mathrm{T}}(z)\mathsf{R}(z)=\mathsf{0}\;,
\end{equation}
and hold for all values of $z$. Now, the fundamental transverse mode in plane $z$ can be written as
\begin{equation}
\label{6u00}
u_{00}(\rho,z)=\sqrt{\frac{k}{\pi\det\mathsf{R}(z)}}\exp\left(-\frac{k\rho^{\mathrm{T}}\mathsf{S}(z)\rho}{2}\right)\;,
\end{equation}
where $\mathsf{S}=-i\mathsf{T}\mathsf{R}^{-1}$. As opposed to $\mathsf{R}$ and $\mathsf{T}$, $\mathsf{S}$ is a $2\times 2$ matrix in the transverse plane and transforms accordingly. It can be checked directly that acting upon $|u_{00}(z)\rangle$ with the lowering operators $\hat{a}_{1}(z)$ and $\hat{a}_{2}(z)$ gives zero. The fundamental mode (\ref{6u00}) is properly normalized and has been constructed such that it solves the paraxial wave equation (\ref{6pwe}) under free propagation. Moreover, one may check that it transforms properly under the transformations of optical elements. The second relation in equation (\ref{6RTprop}) guarantees that $\mathsf{S}$ is symmetric. This is obvious when we multiply the relation from the left with $\big(\mathsf{R}^{\mathrm{T}}\big)^{-1}$, and from the right with $\mathsf{R}^{-1}$. The real and imaginary parts $\mathsf{S}_{\mathrm{r}}$ and $\mathsf{S}_{\mathrm{i}}$ of $\mathsf{S}$ respectively characterize the astigmatism of the intensity and phase patterns. The real part can be written as $\mathsf{S}_{\mathrm{r}}=\big(-i\mathsf{T}\mathsf{R}^{-1}+i(\mathsf{R}^{\dag}\big)^{-1}\mathsf{T}^{\dag})/2$. With the first relation in equation (\ref{6RTprop}) this shows that $\mathsf{R}\mathsf{S}_{\mathsf{r}}\mathsf{R}^{\dag}=\mathsf{1}$. This leads to the identity
\begin{equation}
\label{6Sr}
\mathsf{R}\mathsf{R}^{\dag}=\mathsf{S}_{\mathrm{r}}^{-1}\;,
\end{equation}
which shows that $\mathsf{S}_{\mathrm{r}}$ is positive definite. As a result, the curves of constant intensity in the transverse plane are ellipses. Moreover, the fundamental mode is square-integrable. Depending on the sign of $\det\mathsf{S}_{\mathrm{i}}(z)$ the curves of constant phase in the transverse plane are ellipses, hyperbolas or parallel straight lines. Under free propagation, $\mathsf{S}$ is a slowly varying smooth function of $z$. Optical elements, on the other hand, may instantaneously modify the astigmatism. The astigmatism of both the intensity and the phase patterns is characterized by two widths in mutually perpendicular directions and one angle that specifies the orientation of the curves of constant intensity or phase. The total number of degrees of freedom that specify the astigmatism, and, thereby, the symmetric matrix $\mathsf{S}$, is thus equal to six.

Two of the remaining four degrees of freedom are related to the nature and orientation of the higher-order mode patterns. From equation (\ref{6Sr}), we find that $\mathsf{R}$ can be expressed as $\mathsf{S}_{\mathrm{r}}^{-1/2}\sigma^{\mathrm{T}}$, where $\sigma$ is a unitary $2\times 2$ matrix. Notice that $\mathsf{S}_{\mathrm{r}}$ is real and positive so that $\mathsf{S}_{\mathrm{r}}^{-1/2}$ is well-defined. It is illuminating to rewrite the complex ray vectors $\mu_{1}$ and $\mu_{2}$ as
\begin{equation}
\label{6srsisigmaho}
\left(\!\begin{array}{cc}\mu_{1}&\mu_{2}\end{array}\!\right)=\left(\begin{array}{c}\mathsf{R}\\ \mathsf{T}\end{array}\right)=
\Bigg(\begin{array}{cc}1&0\\-\mathsf{S}_{\mathrm{i}}&1\end{array}\Bigg) \Bigg(\begin{array}{cc}\mathsf{S}_{\mathrm{r}}^{-1/2}&0\\0&\mathsf{S}_{\mathrm{r}}^{1/2}\end{array}\Bigg)
\Bigg(\begin{array}{cc}\sigma^{\mathrm{T}}&0\\0&\sigma^{\mathrm{T}}\end{array}\Bigg)\left(\!\begin{array}{cc}\tilde{\mu}_{x}&\tilde{\mu}_{y}\end{array}\!\right)\;,
\end{equation}
where $\tilde{\mu}_{x}=(1,0,i,0)^{\mathrm{T}}$ and $\tilde{\mu}_{y}=(0,1,0,i)^{\mathrm{T}}$ are the complex ray vectors that correspond to the ladder operators that generate the stationary states of an isotropic harmonic oscillator in two dimensions. The first matrix in the second right-hand-side of this expression (\ref{6srsisigmaho}) is the ray matrix that describes the transformation of a thin astigmatic lens. It imposes the elliptical or hyperbolic wave front of the optical modes on the harmonic-oscillator functions. The second matrix has the form of the ray matrix that describes the deformation of a mode due to refraction. It rescales the modes along two mutually perpendicular transverse directions and accounts for the astigmatism of the intensity patterns. The third matrix involves the complex matrix $\sigma$ and obeys the generalization of equation (\ref{6mgm}) to complex matrices. Since it is complex, however, it is not a ray matrix $\in Sp(4,\mathbb{R})$. In order to clarify its significance, we rewrite equation (\ref{6srsisigmaho}) in terms of the ladder operators, which are conveniently combined in the vector operator $(\hat{a}_{1},\hat{a}_{2})^{\mathrm{T}}$. By using the definition of the ladder operators (\ref{6ladderoperators}) and the Ehrenfest relation (\ref{6ehrenfest}), the transformation in equation (\ref{6srsisigmaho}) can be expressed as
\begin{equation}
\label{6srsisigmaladop}
\left(\begin{array}{c}\hat{a}_{1}\\ \hat{a}_{2}\end{array}\right)=\sqrt{\frac{k}{2}}(\mathsf{R}^{\mathrm{T}}\hat{\theta}-\mathsf{T}^{\mathrm{T}}\hat{\rho})= -i\sqrt{\frac{k}{2}}\sigma\exp\left(-\frac{ik\rho^{\mathrm{T}}\mathsf{S}_{i}\rho}{2}\right)\left(\mathsf{S}_{r}^{1/2}\hat{\rho}+i\mathsf{S}_{r}^{-1/2}\hat{\theta}\right)
\exp\left(\frac{ik\rho^{\mathrm{T}}\mathsf{S}_{i}\rho}{2}\right)\;.
\end{equation}
The linear combination of the position and momentum operators between the brackets takes the form of the lowering-operator vector for an isotropic harmonic oscillator in two dimensions. Again, the $2\times 2$ matrix $\mathsf{S}_{\mathrm{r}}$ accounts for the astigmatism of the intensity patterns by rescaling the ladder operators and, therefore, the modes they generate. The exponential terms take the form of the mode-space transformation for a thin astigmatic lens and impose the curved wave fronts. From right to left, the lowering operators (\ref{6srsisigmaladop}) as well as the corresponding raising operators, first remove the curved wave front, then modify the mode patterns and eventually restore the wave front again. The $2\times 2$ matrix $\sigma$ is a unitary transformation in the space of the lowering operators $\hat{a}_{1}$ and $\hat{a}_{2}$ and transforms accordingly. It arises from the $U(2)$ symmetry of the isotropic harmonic oscillator in two dimensions and accounts for the fact that any, properly normalized, linear combination of bosonic lowering operators yields another bosonic lowering operator. Up to overall phases, to which we come in a moment, this transformation can be parameterized as $\hat{a}_{1}\rightarrow\eta_{1}\hat{a}_{1}+\eta_{2}\hat{a}_{2}$ and $\hat{a}_{2}\rightarrow-\eta^{\ast}_{1}\hat{a}_{1}+ \eta_{2}^{\ast}\hat{a}_{2}$ with $|\eta_{1}|^{2}+|\eta_{2}|^{2}=1$. The two obvious degrees of freedom that are associated with the spinor $\eta=(\eta_{1},\eta_{2})^{\mathrm{T}}$ are the relative amplitude and the relative phase of its components. Analogous to the Poincar\'e sphere for polarization states (or the Bloch sphere for spin-1/2 states), they can be mapped onto a sphere. For reasons that will become clear, this sphere is called the Hermite-Laguerre sphere \cite{Visser04}. Since $\eta_{1}$ and $\eta_{2}$ are spinor components in a linear rather than a circular basis, this mapping takes the following form
\begin{equation}
\label{6etapoincare}
\eta=\left(\begin{array}{c}\eta_{1}\\ \eta_{2}\end{array}\right)=\frac{1}{\sqrt{2}}
\left(\begin{array}{c}
e^{\frac{i\varphi}{2}}\cos\frac{\vartheta}{2}+e^{-i\frac{\varphi}{2}}\sin\frac{\vartheta}{2}\\
-ie^{\frac{i\varphi}{2}}\cos\frac{\vartheta}{2}+ie^{-i\frac{\varphi}{2}}\sin\frac{\vartheta}{2}\end{array}\right)\;,
\end{equation}
where $\vartheta$ and $\varphi$ are the polar and azimuthal angles on the sphere. The mapping is such that the north pole ($\vartheta=0$) corresponds to ladder operators that generate astigmatic Laguerre-Gaussian modes with positive helicity. The south pole ($\vartheta=\pi$) corresponds to Laguerre-Gaussian modes with the opposite helicity while the equator ($\vartheta=\pi/2$) corresponds to Hermite-Gaussian modes. Other values of the polar angle $\vartheta$ correspond to generalized Gaussian modes \cite{Abramochkin04}. The azimuth angle $\varphi$ determines the transverse orientation of the higher-order mode patterns. Since paraxial optical modes are invariant under rotations over $\pi$ in the transverse plane, the mapping in equation (\ref{6etapoincare}) is such that a rotation over $\varphi$ on the sphere corresponds to a rotation of the mode pattern over $\phi=\varphi/2$.

The unitary matrix that describes the ladder operator transformation corresponding to the spinor $\eta$ is constructed as
\begin{equation}
\sigma_{0}(\eta)=\left(\begin{array}{cc}\eta_{1}&\eta_{2}\\-\eta^{\ast}_{2}&\eta^{\ast}_{1}\end{array}\right)\;,
\end{equation}
where the second row is fixed up to a phase factor by the requirement that $\sigma_{0}$ must be unitary. With this convention, the two rows of sigma correspond to antipodal points on the Hermite-Laguerre sphere. Completely fixing the matrix $\sigma\in U(2)$, however, requires four independent degrees of freedom. The remaining two, which are not incorporated in $\eta$, are phase factors. Any matrix $\sigma\in U(2)$ can be written as
\begin{equation}
\label{6sigma0}
\sigma=\left(\begin{array}{cc}e^{i\chi_{1}}&0\\ 0&e^{i\chi_{2}}\end{array}\right)\sigma_{0}(\eta)\;.
\end{equation}
The phase factors $\exp(i\chi_{p})$ correspond to overall phases of the vectors $\mu_{p}$ and, therefore, of the ladder operators (\ref{6ladderoperators}). The vectors $\mu_{p}$ can be written as
\begin{equation}
\label{6nu}
\mu_{p}=e^{i\chi_{p}}\nu_{p}(\mathsf{S},\eta)\;,
\end{equation}
where $p=1,2$ and $\nu_{p}(\mathsf{S},\eta)$ is completely determined by $\mathsf{S}$ and $\eta$ according to (\ref{6srsisigmaho}), $\sigma$ being replaced by $\sigma_{0}(\eta)$. Although the vectors $\nu_{1}$ and $\nu_{2}$ obey symplectic orthonormality conditions (\ref{6muprop}) and are, therefore, not independent, the phases $\chi_{1}$ and $\chi_{2}$ are independent. From equation (\ref{6sigma0}) and the fact that $\mathsf{R}=\mathsf{S}_{\mathrm{r}}^{-1/2}\sigma^{\mathrm{T}}$ it is clear that the argument of $\det \mathsf{R}$ is equal to $\chi_{1}+\chi_{2}$ so that the overall phase of the fundamental mode (\ref{6u00}) is given by $-(\chi_{1}+\chi_{2})/2$. The overall phases of the two raising operators are respectively $-\chi_{1}$ and $-\chi_{2}$, so that the phase factors in the higher order modes $|u_{nm}(z)\rangle$ are given by $\exp(-i\chi_{nm})$ with
\begin{equation}
\label{6chinm}
\chi_{nm}=\left(n+\frac{1}{2}\right)\chi_{1}+\left(m+\frac{1}{2}\right)\chi_{2}\;.
\end{equation}
In a single transverse plane, such overall phase factors do not modify the physical properties of the mode pattern. The evolution of these phases under propagation and optical elements, however, can be measured interferometrically.

The astigmatism of the modes, as characterized by the $2\times 2$ matrix $\mathsf{S}$, can be modified in any desired way by appropriate combinations of the optical elements that we have discussed in section \ref{6Canonical description of paraxial optics}. The degrees of freedom associated with the spinor $\eta$ can be manipulated by mode convertors and image rotators \cite{Beijersbergen93, Galvez03}. Although we shall see that variation of the phase factors $\exp(i\chi_{p})$ is, in general, unavoidable when the other parameters are modified, we show here that it is possible to construct a ray matrix $\in Sp(4,\mathbb{R})$ that solely changes these phase factors. Such a ray matrix is defined by the requirement that
\begin{equation}
M_{\chi}\big(\{\chi_{p}\}\big)\left(\!\begin{array}{cccc}\mu_{1}\;\mu_{2}\;\mu_{1}^{\ast}\;\mu_{2}^{\ast}\end{array}\!\right)=
\left(\!\begin{array}{cccc}e^{i\chi_{1}}\mu_{1}\;e^{i\chi_{2}}\mu_{2}\;e^{-i\chi_{1}}\mu_{1}^{\ast}\;e^{-i\chi_{2}}\mu_{2}^{\ast}\end{array}\!\right)\;,
\end{equation}
so that the vectors $\mu_{p}$ and $\mu_{p}^{\ast}$ are eigenvectors of $M_{\chi}$. The corresponding eigenvalues are unitary. In terms of $\mathsf{R}$ and $\mathsf{T}$ this relation can be expressed as
\begin{equation}
M_{\chi}\big(\{\chi_{p}\}\big)\left(\begin{array}{cc}\mathsf{R}&\mathsf{R}^{\ast}\\ \mathsf{T}&\mathsf{T}^{\ast}\end{array}\right)=\left(\begin{array}{cc}\mathsf{R}&\mathsf{R}^{\ast}\\ \mathsf{T}&\mathsf{T}^{\ast}\end{array}\right)\left(\begin{array}{cc}\mathsf{C}&0\\0&\mathsf{C}^{\ast}\end{array}\right)\;,
\end{equation}
where
\begin{equation}
\label{6matrixchi}
\mathsf{C}=\left(\begin{array}{cc}e^{i\chi_{1}}&0\\0&e^{i\chi_{2}}\end{array}\right)\;.
\end{equation}
By using that
\begin{equation}
\left(\begin{array}{cc}\mathsf{R}&\mathsf{R}^{\ast}\\ \mathsf{T}&\mathsf{T}^{\ast}\end{array}\right)^{-1}=\frac{1}{2i}\left(\begin{array}{cc}-\mathsf{T}^{\dag}&\mathsf{R}^{\dag}\\ \mathsf{T}^{\mathrm{T}}&-\mathsf{R}^{\mathrm{T}}\end{array}\right)\;,
\end{equation}
which follows directly from the identities in equation (\ref{6RTprop}), we find that $M_{\chi}$ can be expressed as
\begin{eqnarray}
\label{6Mchi}
M_{\chi}(\{\chi_{p}\})=\frac{1}{2i}\left(\begin{array}{cc}\mathsf{R}&\mathsf{R}^{\ast}\\ \mathsf{T}&\mathsf{T}^{\ast}\end{array}\right)\left(\begin{array}{cc}\mathsf{C}&0\\0&\mathsf{C}^{\ast}\end{array}\right)\left(\begin{array}{cc}-\mathsf{T}^{\dag}&\mathsf{R}^{\dag}\\ \mathsf{T}^{\mathrm{T}}&-\mathsf{R}^{\mathrm{T}}\end{array}\right)=\qquad\qquad\qquad\qquad\nonumber\\
\frac{1}{2i}\left(
\begin{array}{cc}
-\mathsf{R}\mathsf{C}\mathsf{T}^{\dag}+\mathsf{R}^{\ast}\mathsf{C}^{\ast}\mathsf{T}^{\mathrm{T}}&
\mathsf{R}\mathsf{C}\mathsf{R}^{\dag}-\mathsf{R}^{\ast}\mathsf{C}^{\ast}\mathsf{R}^{\mathrm{T}}\\
-\mathsf{T}\mathsf{C}\mathsf{T}^{\dag}+\mathsf{T}^{\ast}\mathsf{C}^{\ast}\mathsf{T}^{\mathrm{T}}&
\mathsf{T}\mathsf{C}\mathsf{R}^{\dag}-\mathsf{T}^{\ast}\mathsf{C}^{\ast}\mathsf{R}^{\mathrm{T}}
\end{array}\right)\;.
\end{eqnarray}
This ray matrix adds overall phases $\exp(\pm i\chi_{p})$ to the eigenvectors $\mu_{p}$ and $\mu_{p}^{\ast}$. It is real and one may check that it obeys the identity (\ref{6mgm}) so that it is a physical ray matrix $\in Sp(4,\mathbb{R})$.

In this section, we have argued that the number of degrees of freedom associated with the choice of two pairs of ladder operators that generate a basis set of modes in a transverse plane $z$ is equal to the number of generators of $Sp(4,\mathbb{R})$, which is ten. We have shown that six of those are related to the astigmatism of the modes as characterized by the complex and symmetric $2\times 2$ matrix $\mathsf{S}$. Two of the other four are angles on the Hermite-Laguerre sphere that characterize a spinor $\eta$, which determines the nature and orientation of the higher-order modes. The remaining two are overall phases of the ladder operators. All these degrees of freedom can be manipulated in any desired way by choosing a suitable ray matrix $\in Sp(4,\mathbb{R})$.

\subsection{Gouy phase}
In the limiting case of non-astigmatic modes that propagate through an isotropic optical system the $2\times 2$ matrix $\mathsf{S}$ is a symmetric matrix with degenerate eigenvalues so that it can be considered a scalar $s=s_{\mathrm{r}}+is_{\mathrm{i}}$. If we choose $\sigma_{0}=1$, the higher-order modes are Hermite-Gaussian. In that case, the complex ray vectors are given by $\mu_{1}=(r,0,t,0)^{\mathrm{T}}$ and $\mu_{2}=(0,r,0,t)^{\mathrm{T}}$, with $r,t\in\mathbb{C}$. The symplectic normalization condition (\ref{6muprop}) implies that $r^{\ast}t-t^{\ast}r=2i$. The real part $s_{\mathrm{r}}$ of $s=-it/r$ determines the beam width $w=\sqrt{2/(ks_{r})}$ of the fundamental mode while the imaginary part $s_{\mathrm{i}}$ fixes the radius of curvature of its wave fronts according to $R=1/s_{\mathrm{i}}$. Under free propagation over a distance $z$, the vectors $\mu_{1}$ and $\mu_{2}$ transform according to
\begin{equation}
\label{6mu12isotropic}
\mu_{1}(z)=\left(\begin{array}{c}r+zt\\0\\t\\0\end{array}\right)\quad\mathrm{and}\quad\mu_{2}(z)= \left(\begin{array}{c}0\\r+zt\\0\\t\end{array}\right)\;.
\end{equation}
The parameters $r$, $t$ and $s$ remain scalar and free propagation does not introduce an overall phase difference between $\mu_{1}$ and $\mu_{2}$ so that $\eta$, or, equivalently $\sigma_{0}$, is independent of $z$. Without loss of generality we can choose $z=0$ to coincide with the focal plane of the mode, which implies that $s\in\mathbb{R}$ so that $r^{\ast}t=-t^{\ast}r=i$. Since $s_{\mathrm{r}}$, and, therefore, $\mathsf{R}=\sigma_{0}s_{\mathrm{r}}$ cannot pick up a phase, we find that
\begin{equation}
\chi(z)-\chi(0)=\arg\left(\frac{r+zt}{r}\right)=\arctan\left(\frac{tz}{r}\right)=\arctan\left(\frac{z}{z_{\mathrm{R}}}\right)\;,
\end{equation}
where $z_{\mathrm{R}}=ir/t$ is the Rayleigh range. This is the well-known Gouy phase for a Gaussian mode \cite{Siegman}. Since the vectors $\mu_{1}$ and $\mu_{2}$ pick up an overall phase $\chi(z)$, the raising operators pick up a phase $-\chi(z)$. The phase shift of the higher-order modes (\ref{6modes}) is then given by $\exp(-i(n+m+1)\chi)$ and depends on the total mode number $N=n+m$ only. As a result of this degeneracy, the same expression holds in the non-astigmatic case with $\sigma_{0}\neq 1$. In that case, it is still true that the components of $\eta$ are independent of $z$.

Generalization to astigmatic modes is straightforward only if the modes have simple astigmatism and if the orientation of the higher-order mode patterns is aligned along the astigmatism of the fundamental mode. In that case, the vectors $\mu_{p}$ pick up different Gouy phases and the components of $\eta$ are independent of $z$. This is not true in the case of non-astigmatic modes that propagate through an optical set-up with simple astigmatism \cite{Beijersbergen93}. In the more general case of modes with general astigmatism that propagate through an arbitrary set-up of paraxial optical elements, the $z$ dependence of $\mathsf{S}$ depends on $\eta$ and vice versa \cite{Visser04}. In this case no simple analytical expressions of the Gouy phases can be derived. The phase in equation (\ref{6chinm}) may be viewed as the ultimate generalization of the Gouy phase within paraxial wave optics.

\section{The geometric interpretation of the variation of the phases $\chi_{nm}$}
\label{6The geometric origin of variation of the phases}
\subsection{Evolution of the phases $\chi_{nm}$}
In this section we show that variation of the phase differences $\chi_{p}$ between $\mu_{p}$ and $\nu_{p}$ (\ref{6nu}) is, in general, unavoidable under (a sequence of) mode transformations that modify the degrees of freedom associated with $\mathsf{S}$ and $\eta$. From the discussion in the previous section it is clear that the generalized Gouy phases were defined such that they vary only under transformations that involve free propagation. However, it is convenient to formulate the description of mode transformations that give rise to phase shifts in a slightly more general way.

Suppose that the unitary state-space transformation that describes (a part of) a trajectory through the parameter space is given by $\hat{U}(\zeta)=\exp(-i\hat{T}\zeta)$, where $\hat{T}$ is a (linear combination of the) generator(s) defined in equation (\ref{6generators}) and $\zeta$ is a real parameter that parameterizes the trajectory. In this case, the $\zeta$ dependent ladder operators (\ref{6ladoptrans}) obey the anti-Heisenberg equation of motion
\begin{equation}
\label{6antiheisenberg}
\Big[\hat{a}^{\left(\dag\right)}(\zeta),\hat{T}\Big]=-i\frac{\partial \hat{a}^{\left(\dag\right)}}{\partial \zeta}\;.
\end{equation}
In terms of the complex ray vectors $\mu_{p}(\zeta)$ and the ray matrix $M(\zeta)=\exp(-J\zeta)$ that corresponds to $\hat{U}(\zeta)$ according to relation (\ref{6ehrenfest}), this equation of motion takes the form of a symplectic Schr\"odinger equation and can be expressed as
\begin{equation}
\label{6eqmoJ1}
\frac{\partial\mu_{p}}{\partial \zeta}=-J\mu_{p}(\zeta)\;.
\end{equation}
Substitution of $\mu_{p}(\zeta)=\exp(i\chi_{p})\nu_{p}(\zeta)$ yields after dividing by $\exp(i\chi_{p})$
\begin{equation}
i\frac{\partial\chi_{p}}{\partial \zeta}\nu_{p}(\zeta)+\frac{\partial\nu_{p}}{\partial \zeta}=-J\nu_{p}(\zeta)\;.
\end{equation}
By multiplying from the left with $\nu_{p}^{\dag}G$, using the normalization condition $\nu_{p}^{\dag}G\nu_{p}=2i$ and rearranging the terms we find that
\begin{equation}
\label{6dchictau}
\frac{\partial\chi_{p}}{\partial \zeta}=\frac{1}{2}\left\{\nu_{p}^{\dag}GJ\nu_{p}+\nu_{p}^{\dag}G\frac{\partial\nu_{p}}{\partial \zeta}\right\}\;.
\end{equation}
The generator $J$ represents a conserved quantity. Hence, the first term between the curly brackets does not depend on the parameter $\zeta$ and the above equation (\ref{6dchictau}) can be integrated to obtain
\begin{equation}
\label{6dyngeomphase}
\chi_{p}(\zeta)=\frac{1}{2}\left\{\left(\nu_{p}^{\dag}GJ\nu_{p}\right)\zeta+\int_{0}^{\zeta}d\zeta'\nu_{p}^{\dag}G\frac{\partial\nu_{p}}{\partial \zeta'}\right\}\;.
\end{equation}
The first term between the curly brackets constitutes a dynamical contribution to the phase shift and arises from the fact that $J$ corresponds to a constant of motion. The second term, on the other hand, relates to the geometry of the complex ray space and is the natural generalization of Berry's geometric phase to this case. In the next section, we derive an equivalent expression from which the geometric significance of the phase shifts (\ref{6dyngeomphase}) is more obvious.

\subsection{Analogy with the Aharonov-Bohm effect}
As is already shown in Berry's original paper \cite{Berry84}, the Aharonov-Bohm effect in quantum electrodynamics may be viewed as an example of a geometric phase. Conversely, it is well-known that any geometric phase shift may be interpreted as being the result of the coupling to a (fictitious) gauge field, or vector potential, $\vec{A}$. The corresponding ``magnetic'' field $F_{\alpha\beta}=\partial_{\alpha}A_{\beta}-\partial_{\beta}A_{\alpha}$, where the indices $\alpha$ and $\beta$ run over the vector components, is called the Berry curvature.

In the present case, the underlying local gauge invariance relates to the fact that the physical properties of the mode fields (\ref{6modes}), for instance those in equation (\ref{6unmTJ}), are locally not affected by transformations of the following type
\begin{equation}
\label{6gaugetransform}
\mu_{p}\rightarrow e^{i\psi_{p}\left(\vec{\mathcal{R}}\right)}\mu_{p}\;,
\end{equation}
where $p=1,2$ and $\psi_{p}$ are real phases. This property constitutes a local $U(1)\otimes U(1)$ gauge invariance. The ray matrix $\in Sp(4,\mathbb{R})$ that describes such gauge transformations (\ref{6gaugetransform}) figures in equation (\ref{6Mchi}). As shown in appendix \ref{6Mode-space operators corresponding to the Noether charges}, the two corresponding real generators $J_{\chi_{p}}$ can be constructed from the eigenvectors $\mu_{p}$. The vector $\mu_{1}$ is an eigenvector of $J_{\chi_{1}}$ with eigenvalue $-i$. Since $J_{\chi_{1}}$ is real, the complex conjugate vector $\mu_{1}^{\ast}$ is an eigenvector of $J_{\chi_{p}}$ with eigenvalue $i$. Moreover, $J_{\chi_{1}}\mu_{2}=J_{\chi_{1}}\mu^{\ast}_{2}=0$. Similarly, $\mu_{2}$ and $\mu_{2}^{\ast}$ are eigenvectors of $J_{\chi_{2}}$ with eigenvalues $-i$ and $i$, and $J_{\chi_{2}}\mu_{1}=J_{\chi_{2}}\mu^{\ast}_{1}=0$. Since invariance under the gauge transformation (\ref{6gaugetransform}) is a local and continuous symmetry, it gives rise to conserved Noether charges. The gauge transformations are generated by two different generators, hence there are two Noether charges, which can be expressed as $\nu_{p}^{\dag}GJ_{\chi_{p}}\nu_{p}/2=1$, where the factor $1/2$ arises from the fact that a symplectic vector space is a joint space of position and momentum and where we have used that $J_{\chi_{p}}\nu_{p}=-i$ and $\nu_{p}^{\dag}G\nu_{p}=2i$. In appendix \ref{6Mode-space operators corresponding to the Noether charges}, we prove that the corresponding state-space generators $\hat{T}_{\chi_{p}}$ can be expressed as $\Big(\hat{a}^{\dag}_{p}\hat{a}_{p}+\hat{a}_{p}\hat{a}^{\dag}_{p}\Big)/2$ so that the charges of a mode (\ref{6modes}) are given by $\langle u_{nm}|\hat{T}_{\chi_{1}}|u_{nm}\rangle=(n+1/2)$ and $\langle u_{nm}|\hat{T}_{\chi_{2}}|u_{nm}\rangle=(m+1/2)$. Since the gauge transformation in equation (\ref{6Mchi}) is constructed from the eigenvectors $\mu_{p}$, it varies throughout the parameter space. As a result, the generators $\hat{T}_{\chi_{p}}$ can be constructed only locally and vary through the parameter space according to the ladder-operator transformation given in equation (\ref{6ladoptrans}). However, since the modes also vary, it follows that the Noether charges $(n+1/2)$ and $(m+1/2)$ of the modes $|u_{nm}\rangle$ are globally conserved.

In order to derive an expression of the geometric phase shifts (\ref{6dyngeomphase}) that relates them to the underlying gauge symmetry (\ref{6gaugetransform}), it is convenient to rewrite the equations of motion of the vectors $\mu_{p}$ (\ref{6eqmoJ1}) as
\begin{equation}
\label{6eqmoJ2}
\left(\vec{\nabla}_{\vec{\mathcal{R}}}\mu_{p}\right)\cdot\frac{\partial\vec{\mathcal{R}}}{\partial \zeta}=-J\mu_{p}\;,
\end{equation}
where $\vec{\mathcal{R}}=(\mathcal{R}_{1},\mathcal{R}_{2},...)^{\mathrm{T}}$ is a vector of eight parameters that characterize the degrees of freedom associated with the matrix $\mathsf{S}$ and spinor $\eta$ and $\vec{\nabla}_{\vec{\mathcal{R}}}=(\partial/\partial\mathcal{R}_{1},\partial/\partial\mathcal{R}_{2},...) ^{\mathrm{T}}$ is the corresponding differential operator. The equations of motion (\ref{6eqmoJ2}) are obviously not invariant under the gauge transformations (\ref{6gaugetransform}). Imposing gauge invariance yields the modified equations of motion
\begin{equation}
\label{6eqmogaugefield}
\Big(\left(\vec{\nabla}_{\vec{\mathcal{R}}}+i\vec{A}_{p}\right)\nu_{p}\Big)\cdot\frac{\partial\vec{\mathcal{R}}}{\partial \zeta}=-J\nu_{p}\;,
\end{equation}
where the gauge fields $\vec{A}_{p}$ are vector fields in the parameter space of $\vec{\mathcal{R}}$ that are defined by their transformation property under the gauge transformations (\ref{6gaugetransform})
\begin{equation}
\label{6gaugefieldtransprop}
\vec{A}_{p}\rightarrow\vec{A}_{p}-\vec{\nabla}_{\vec{\mathcal{R}}}\psi_{p}\;.
\end{equation}
With these transformation properties, the equation of motion (\ref{6eqmogaugefield}) is manifestly invariant under the gauge transformations (\ref{6gaugetransform}). The general solution of this equation (\ref{6eqmogaugefield}) can be expressed as
\begin{equation}
\label{6geomphasegaugefield}
\nu_{p}=\mu_{p}e^{-i\int_{\mathcal{C}}\vec{A}_{p}\cdot d\vec{\mathcal{R}}}\;,
\end{equation}
where $\mathcal{C}$ is a trajectory $\vec{\mathcal{R}}(\zeta)$ and $\mu_{p}$ solves the equation of motion without the gauge field (\ref{6eqmoJ2}). In full analogy with the Aharonov-Bohm effect, this shows that the phase difference between $\mu_{p}$ and $\nu_{p}$ is due to the fact that the latter is coupled to the gauge field $\vec{A}_{p}$. Since we have defined the vectors $\mu_{p}$ so as to include the appropriate geometric-phase factor while they are not coupled to the gauge fields, the coupling of $\nu_{p}$ to the gauge fields removes the geometric phase rather than introducing it. The geometric origin of the phases is evident in that they are determined only by the trajectory $\mathcal{C}$ and do not depend on the velocity $\partial\vec{\mathcal{R}}/\partial \zeta$. By using equation (\ref{6eqmogaugefield}) they can be expressed as
\begin{equation}
\chi_{p}=\int_{\mathcal{C}}\vec{A}_{p}\cdot d\vec{\mathcal{R}}=\frac{1}{2}\int_{0}^{\zeta}d\zeta'\left\{\nu_{p}GJ\nu_{p}+\left(\nu_{p}^{\dag}G\vec{\nabla}_{\mathcal{R}}\nu_{p}\right) \cdot\frac{\partial\vec{R}}{\partial \zeta}\right\}\;,
\end{equation}
which is in clear accordance with equation (\ref{6dyngeomphase}).

In analogy with the Aharonov-Bohm effect, the Noether charges $\nu_{p}^{\dag}GJ_{\chi_{p}}\nu_{p}/2=1$ determine the strength of the coupling of the vectors $\mu_{p}$ to the gauge fields $\vec{A}_{p}$. This is consistent with the fact that the vectors $\nu_{p}$ pick up phases $\chi_{p}$. The Noether charges of the modes (\ref{6modes}), however, are equal to $n+1/2$ and $m+1/2$ and depend on the mode numbers $n$ and $m$. As a result, the modes $|u_{nm}\rangle$ couple differently to the (corresponding state-space) gauge fields and, therefore, experience different phase shifts. This is in obvious agreement with equation (\ref{6chinm}).

The Noether currents $(\nu_{p}^{\dag}GJ_{\chi_{p}}\nu_{p}/2)\partial\vec{\mathcal{R}}/\partial \zeta=\partial\vec{\mathcal{R}}/\partial \zeta$ are uniform throughout the parameter space of $\vec{\mathcal{R}}$. It follows that the corresponding Berry curvatures do not possess any non-trivial dynamics. Attributing the generalized Gouy phases $\chi_{p}$ to coupling to gauge fields $\vec{A}_{p}$, which do not have any dynamical properties in their own rights, may seem a bit tautological. On the other hand, the analysis discussed here shows that the structure that underlies the generalized Gouy phase shifts (\ref{6chinm}) is that of a gauge theory. In this picture, the appearance of phase shifts under propagation through an optical set-up is the unavoidable consequence of the $U(1)\times U(1)$ gauge invariance of the dynamics of paraxial optical modes, or, equivalently, of the fact that the mode charges $n+1/2$ and $m+1/2$ are conserved under state-space transformations $\in Mp(4,\mathbb{R})$.

\section{Concluding remarks}
We have explored the parameter space that is associated with the choice of a complete and orthonormal set of paraxial optical modes in the transverse plane. Modes are defined as solutions of the paraxial wave equation (\ref{6pwe}) that are fully characterized by a set of mode parameters whose variation through a paraxial optical set-up is described by the $4\times 4$ ray matrix $M(z)$, which describes the transformation of a ray $r=(\rho,\theta)^{\mathrm{T}}$ from the $z=0$ input plane of the set-up to the transverse plane $z$. Complete sets of transverse modes can be obtained from two pairs of bosonic ladder operators. The ladder operators are fully specified by two complex ray vectors $\mu_{p}$ with $p=1,2$, which characterize the mode parameters. Their variation through an optical set-up, and, thereby, the variation of the ladder operators, can conveniently be expressed in terms of $M(z)$. We have argued that there is a one-to-one correspondence between the algebraic properties of the ladder operators and the defining properties of a physical ray matrix $\in Sp(4,\mathbb{R})$, i.e., that it is real and obeys the identity (\ref{6mgm}). It follows that all sets of modes can be expressed in terms of two pairs of ladder operators and, moreover, that the freedom in choosing a set of modes is equivalent to the choice of an arbitrary ray matrix $M_{0}\in Sp(4,\mathbb{R})$. Since $Sp(4,\mathbb{R})$ is a ten-parameter Lie group, the number of free parameters associated with this choice is equal to ten. A possible physical characterization of these degrees of freedom involves a symmetric $2\times 2$ matrix $\mathsf{S}$, which characterizes the astigmatism of the phase and intensity patterns of the fundamental mode, and a spinor $\eta$, which specifies the nature and orientation of the higher-order modes. The matrix $\mathsf{S}$ is fully specified by six parameters while characterization of $\eta$ requires two independent parameters, which can be mapped on a Poincar\'e sphere. The remaining two degrees of freedom are overall phases of the ladder operators. They do not modify the physical properties of the modes in a given transverse plane $z$. Their variation through an optical set-up, however, gives rise to a generalized Gouy phase shift of the modes, which can be measured interferometrically. We have shown that both contributions to the variation of the overall phases through an optical set-up, as described by equation (\ref{6dyngeomphase}), are geometric in that they are fully determined by the trajectory $\vec{\mathcal{R}}(\zeta)$ and do not depend on the velocity $\partial\vec{\mathcal{R}}/\partial \zeta$. However, only the second contribution in equation (\ref{6dyngeomphase}) relates to the geometry of the parameter space.

It is noteworthy that the overall phases $\chi_{p}$ of the vectors $\mu_{p}$ are in general only unambiguously defined in case of a closed trajectory. In particular, in the propagation-direction representation, the astigmatism of the fundamental mode $\tilde{u}_{00}(\theta,z)$ is fully specified by the symmetric matrix $\mathsf{V}=\mathsf{S}^{-1}$. Analogous to the discussion in section \ref{6Basis sets of paraxial modes}, the remaining degrees of freedom can be characterized by a unitary $2\times 2$ matrix $\upsilon$, which is defined such that $\mathsf{T}=\mathsf{V}_{\mathrm{r}}^{-1/2} \upsilon^{\mathrm{T}}$. It follows that $\upsilon$ and $\sigma$ are related by $\sigma=-i\upsilon\mathsf{V}_{\mathrm{r}}^{-1/2}\mathsf{V}\mathsf{S}_{\mathrm{r}}^{1/2}$. In general $\det\left(\mathsf{V}_{\mathrm{r}}^{-1/2}\mathsf{V}\mathsf{S}_{\mathrm{r}}^{1/2}\right)\neq 1$ so that defining $\sigma=\mathsf{C}\sigma_{0}$ and  $\upsilon=\mathsf{C}'\upsilon_{0}$ such that $\sigma_{0}$ and $\upsilon_{0}$ have unit determinants, requires different phase matrices $\mathsf{C}\neq\mathsf{C}'$. The phase shift along a closed trajectory, however, does not depend on the phase convention used. In the limiting case of transformations of non-astigmatic modes in their focal planes, i.e., when $\mathsf{S}$ and $\mathsf{V}$ can be considered real scalars, the phases are also unambiguously defined along an open trajectory. The central results presented here, and, in particular, the formulation in section \ref{6The geometric origin of variation of the phases} are, of course, independent of the phase convention chosen.

We have shown that the symplectic group of ladder-operator transformations $Sp(4,\mathbb{R})$ corresponds to the metaplectic group $Mp(4)$ of unitary transformations on the Hilbert space of state vectors $|u\rangle$. The metaplectic group constitutes a subgroup of the set of all possible unitary transformations. This accounts for the fact that only specific linear combinations of paraxial optical modes are modes as well, i.e., are fully characterized by a set of parameters whose variation through a paraxial optical set-up is fully described by the ray matrix $M(z)$. Each combination $(n,m)$ of the transverse mode indices gives rise to a subspace of the Hilbert space of transverse states of the field, which is closed under metaplectic transformations. These subspaces are all isomorphic to the symplectic manifold underlying the ladder operators.

Analogous to the Aharonov-Bohm effect, the variation of the geometric phases may also be derived from a local gauge invariance of the ladder-operator dynamics. This allows for deriving an expression of the geometric phase shifts in terms of the gauge fields that arise from this symmetry, which explicitly reveals their geometric origin. Moreover, it provides a deep, purely geometric, understanding of the factors $n+1/2$ and $m+1/2$ in the expression of the generalized Gouy phase shift of the modes (\ref{6dyngeomphase}). These factors appear as conserved Noether charges, which arise from the underlying gauge symmetry and determine the strength of the coupling of the modes to the (corresponding state-space) gauge fields. Notice that the above-mentioned subspaces of modes with transverse mode indices $n$ and $m$ are all uniquely characterized by these coupling strengths.

Although we have focused on the optical case, the mathematical structure that underlies the ladder-operator method and the phase shifts that arise from the geometry underlying the ladder operators are more general. The ray space $(\rho,\theta)$ is a phase space in the mathematical sense and the operator description of paraxial wave optics that we have discussed in section \ref{6Canonical description of paraxial optics} may be viewed as a formally quantized (wavized) description of rays. Although the interpretation is different, all this is in full analogy with the quantization of classical mechanics to obtain quantum mechanics. As a result, the methods and results of this chapter can be applied to the quantum-mechanical description of wave packets. The only restriction for the ladder-operator approach to apply is that the state-space generators (or Hamiltonian in the quantum language) are quadratic in the canonical operators. The methods and results in this chapter have been formulated such that it is evident how they can be generalized to account for more independent spatial dimensions. In the general case of $D$ dimensions, the number of generators of $Mp(2D)$ and $Sp(2D,\mathbb{R})$ is equal to $2D^{2}+D$, $D^2+D$ of which are associated with a $D\times D$ symmetric matrix that generalizes $\mathsf{S}$. The remaining $D^2$ parameters specify a unitary matrix $\in U(D)$, which generalizes $\sigma$, and corresponds to the choice of $D$ orthonormal $D-$component spinors and $D$ overall phase factors. The variation of the phases under propagation (evolution) have a geometric interpretation in terms of the other degrees of freedom.

\newpage

\section*{Appendices}
\begin{subappendices}
\section{The ray-space generators $J_{j}$}
\label{6The ray-space generators}
In this appendix we give explicit expressions of the ray-space generators $J_{j}$. They are defined by equation (\ref{6homomorphism}) and correspond to the state-space generators $\hat{T}_{j}$ as defined in equation (\ref{6generators}). They are given by
\begin{eqnarray}
&J_{1}=\frac{2}{k}\left(\begin{array}{cccc}0&0&0&0\\0&0&0&0\\1&0&0&0\\0&0&0&0\end{array}\right)\quad
J_{2}=\frac{2}{k}\left(\begin{array}{cccc}0&0&0&0\\0&0&0&0\\0&0&0&0\\0&1&0&0\end{array}\right)\quad
J_{3}=\frac{1}{k}\left(\begin{array}{cccc}0&0&0&0\\0&0&0&0\\0&1&0&0\\1&0&0&0\end{array}\right)&\nonumber\\
&J_{4}=\left(\begin{array}{cccc}-1&0&0&0\\0&0&0&0\\0&0&1&0\\0&0&0&0\end{array}\right)\quad
J_{5}=\left(\begin{array}{cccc}0&0&0&0\\0&-1&0&0\\0&0&0&0\\0&0&0&1\end{array}\right)&\nonumber\\
&J_{6}=\left(\begin{array}{cccc}0&0&0&0\\-1&0&0&0\\0&0&0&1\\0&0&0&0\end{array}\right)\quad
J_{7}=\left(\begin{array}{cccc}0&-1&0&0\\0&0&0&0\\0&0&0&0\\0&0&1&0\end{array}\right)&\nonumber\\
&J_{8}=k\left(\begin{array}{cccc}0&0&0&-1\\0&0&-1&0\\0&0&0&0\\0&0&0&0\end{array}\right)\quad
J_{9}=2k\left(\begin{array}{cccc}0&0&-1&0\\0&0&0&0\\0&0&0&0\\0&0&0&0\end{array}\right)\quad
J_{10}=2k\left(\begin{array}{cccc}0&0&0&0\\0&0&0&-1\\0&0&0&0\\0&0&0&0\end{array}\right)
\end{eqnarray}

\section{Expectation values of the generators $\hat{T}_{j}$}
\label{6Expectation values of the generators}
This appendix is devoted to a proof of equation (\ref{6unmTJ}), which expresses the expectation values $\langle u_{nm}|\hat{T}_{j}|u_{nm}\rangle$ of the generators $\hat{T}_{j}$ in equation (\ref{6generators}) in terms of the corresponding ray-space generators $\hat{J}_{j}$ as defined by equation (\ref{6homomorphism}). We prove this by mathematical induction. The special cases $\langle u_{00}|\hat{T}_{j}|u_{00}\rangle$ involve Gaussian standard integrals and can be proven explicitly. A formal proof by mathematical induction thus requires showing that the identity (\ref{6unmTJ}) holds for modes $|u_{n+1m}\rangle$ and $|u_{nm+1}\rangle$ if it holds for $|u_{nm}\rangle$. In order to prove this, we notice that
\begin{equation}
\langle u_{n+1m}|\hat{T}_{j}|u_{n+1m}\rangle=\frac{1}{n+1}\langle u_{nm}|\hat{a}_{1}\hat{T}_{j}\hat{a}_{1}^{\dag}|u_{nm}\rangle\;.
\end{equation}
Using that
\begin{equation}
\left[\hat{T}_{j},\hat{a}_{p}\right]=\sqrt{\frac{k}{2}}\left(\mu_{p}^{\mathrm{T}}G\hat{T}_{j}\hat{\scripty{r}}-\mu_{p}^{\mathrm{T}}G\hat{\scripty{r}}\hat{T}_{j}\right)=
\sqrt{\frac{k}{2}}\mu_{p}^{\mathrm{T}}G\left[\hat{T}_{j},\hat{\scripty{r}}\right]=i\sqrt{\frac{k}{2}}\mu_{p}^{\mathrm{T}}GJ_{j}\hat{\scripty{r}}\;,
\end{equation}
this can be rewritten as
\begin{eqnarray}
\left(\frac{1}{n+1}\right)\langle u_{nm}|\left(\hat{T}_{j}\hat{a}_{1}-i\sqrt{\frac{k}{2}}\mu_{1}^{\mathrm{T}}GJ_{j}\hat{\scripty{r}}\right)\hat{a}_{1}^{\dag}|u_{nm}\rangle= \qquad\qquad\qquad\qquad\qquad\qquad\nonumber\\
\langle u_{nm}|\hat{T}_{j}|u_{nm}\rangle-\left(\frac{i}{n+1}\right)\sqrt{\frac{k}{2}}\mu_{1}^{\mathrm{T}}GJ_{j}\langle u_{nm}|\hat{\scripty{r}}\hat{a}_{1}^{\dag}|u_{nm}\rangle\;.
\end{eqnarray}
The analogous result may be derived for $|u_{nm+1}\rangle$ and proving equation (\ref{6unmTJ}) thus boils down to proving that
\begin{equation}
-\left(\frac{i}{n+1}\right)\sqrt{\frac{k}{2}}\mu_{p}^{\mathrm{T}}GJ_{j}\langle u_{nm}|\hat{\scripty{r}}\hat{a}_{p}^{\dag}|u_{nm}\rangle=\frac{1}{2}\mu_{p}^{\dag}GJ_{j}\mu_{p}
=\frac{1}{2}\left(\mu_{p}^{\dag}GJ_{j}\mu_{p}\right)^{\mathrm{T}}=\frac{1}{2}\mu_{p}^{T}GJ_{j}\mu_{p}^{\ast}\;,
\end{equation}
where we used that $G^{\mathrm{T}}=-G$ and that $J^{\mathrm{T}}G=-GJ$. This expression can be rewritten as
\begin{equation}
\label{6unmraunm}
\langle u_{nm}|\hat{\scripty{r}}\hat{a}_{p}^{\dag}|u_{nm}\rangle=i(n+1)\sqrt{\frac{1}{2k}}\mu_{p}^{\ast}\;,
\end{equation}
which we also prove by mathematical induction. Again, the special case of $|u_{00}\rangle$ can be checked explicitly. In order to prove that it is true for $|u_{n+1m}\rangle$ and $|u_{nm+1}\rangle$, we use that
\begin{equation}
\left[\hat{\scripty{r}},\hat{a}_{p}^{\dag}\right]=\hat{\scripty{r}}\left(\sqrt{\frac{k}{2}}\mu_{p}^{\dag}G\hat{\scripty{r}}\right)- \left(\sqrt{\frac{k}{2}}\mu_{p}^{\dag}G\hat{\scripty{r}}\right)\hat{\scripty{r}}=
\sqrt{\frac{k}{2}}\left[\hat{\scripty{r}},r_{p}^{\ast}\hat{\theta}-t_{p}^{\ast}\hat{\rho}\right]=i\sqrt{\frac{1}{2k}}\mu_{p}^{\ast}
\end{equation}
and find
\begin{eqnarray}
\langle u_{n+1m}|\hat{\scripty{r}}\hat{a}_{1}^{\dag}|u_{n+1m}\rangle=\left(\frac{1}{n+1}\right)\langle u_{nm}|\hat{a}_{1}\hat{\scripty{r}}\hat{a}_{1}^{\dag}\hat{a}_{1}^{\dag}|u_{nm}\rangle=\qquad\qquad\qquad\qquad\qquad\qquad\nonumber\\
\left(\frac{1}{n+1}\right)\langle u_{nm}|\hat{a}_{1}\left(\hat{a}_{1}^{\dag}\hat{\scripty{r}}+i\sqrt{\frac{1}{2k}}\mu_{1}^{\ast}\right)\hat{a}_{1}^{\dag}|u_{nm}\rangle
=\langle u_{nm}|\hat{\scripty{r}}\hat{a}_{1}^{\dag}|u_{nm}\rangle+i\sqrt{\frac{1}{2k}}\mu_{1}^{\ast}\;.
\end{eqnarray}
The analogous result may be derived for $|u_{nm+1}\rangle$. This completes the proof of equation (\ref{6unmraunm}) and, thereby, of equation (\ref{6unmTJ}).

\section{Mode-space operators corresponding to the Noether charges}
\label{6Mode-space operators corresponding to the Noether charges}
In this appendix we construct both the ray-space and the corresponding state-space generators of the $U(1)\otimes U(1)$ gauge transformations. The ray matrix that describes such transformations is given by equation (\ref{6Mchi}). To first order in the phases $\chi_{1}$ and $\chi_{2}$ the matrix $\mathsf{C}$ (\ref{6matrixchi}) is given by
\begin{equation}
\mathsf{C}=\left(\begin{array}{cc}1+i\chi_{1}&0\\0&1+i\chi_{2}\end{array}\right)=\left(\begin{array}{cc}1&0\\0&1\end{array}\right) +\chi_{1}\left(\begin{array}{cc}i&0\\0&0\end{array}\right)+\chi_{2}\left(\begin{array}{cc}0&0\\0&i\end{array}\right)\;.
\end{equation}
Substitution in equation (\ref{6Mchi}) then gives
\begin{eqnarray}
M_{\chi}\big(\{\chi_{p}\}\big)=1+\frac{\chi_{1}}{2}
\left(\begin{array}{cc}-r_{1}t_{1}^{\dag}-r_{1}^{\ast}t_{1}^{\mathrm{T}}&r_{1}r_{1}^{\dag}+r_{1}^{\ast}r_{1}^{\mathrm{T}}\\
-t_{1}t_{1}^{\dag}-t_{1}^{\ast}t_{1}^{\mathrm{T}}&t_{1}r_{1}^{\dag}+t_{1}^{\ast}r_{1}^{\mathrm{T}}\end{array}\right)
+\qquad\qquad\qquad\qquad\nonumber\\
\frac{\chi_{2}}{2}\left(\begin{array}{cc}-r_{2}t_{2}^{\dag}-r_{2}^{\ast}t_{2}^{\mathrm{T}}&r_{2}r_{2}^{\dag}+r_{2}^{\ast}r_{2}^{\mathrm{T}}\\
-t_{2}t_{2}^{\dag}-t_{2}^{\ast}t_{2}^{\mathrm{T}}&t_{2}r_{2}^{\dag}+t_{2}^{\ast}r_{2}^{\mathrm{T}}\end{array}\right)\;,
\end{eqnarray}
where $r_{1}t_{1}^{\dag}=r_{1}\otimes t_{1}^{\dag}$ etcetera are direct vector products. From $M_{\chi}\big(\{\chi_{p}\}\big)=\exp\left(-\chi_{p}J_{\chi_{p}}\right)\simeq 1-\chi_{p}J_{\chi_{p}}$, we find that
\begin{equation}
J_{\chi_{p}}=\frac{1}{2}\left(\begin{array}{cc}r_{p}t_{p}^{\dag}+r_{p}^{\ast}t_{p}^{\mathrm{T}}&-r_{p}r_{p}^{\dag}-r_{p}^{\ast}r_{p}^{\mathrm{T}}\\
t_{p}t_{p}^{\dag}+t_{p}^{\ast}t_{p}^{\mathrm{T}}&-t_{p}r_{p}^{\dag}-t_{p}^{\ast}r_{p}^{\mathrm{T}}\end{array}\right)
\end{equation}
where $p=1,2$. These generators are $4\times 4$ matrices in the ray space. By carefully inspecting the form of the direct products and the structure of the generators $J_{j}$ as given in appendix \ref{6The ray-space generators} we find that
\begin{eqnarray}
\hat{T}_{\chi_{p}}=-\frac{k}{4}\Big\{
r_{p}^{\mathrm{T}}\hat{\theta}\hat{\rho}^{\mathrm{T}}t_{p}^{\ast}+r_{p}^{\dag}\hat{\theta}\hat{\rho}^{\mathrm{T}}t_{p}^{\mathrm{T}}
-r_{p}^{\mathrm{T}}\hat{\theta}\hat{\theta}^{\mathrm{T}}r_{p}^{\ast}-r_{p}^{\dag}\hat{\theta}\hat{\theta}^{\mathrm{T}}r_{p}^{\mathrm{T}}+\qquad\qquad\qquad\quad\nonumber\\ t_{p}^{\mathrm{T}}\hat{\rho}\hat{\theta}^{\mathrm{T}}r_{p}^{\ast}+t_{p}^{\dag}\hat{\rho}\hat{\theta}^{\mathrm{T}}r_{p}^{\mathrm{T}}
-t_{p}^{\mathrm{T}}\hat{\rho}\hat{\rho}^{\mathrm{T}}t_{p}^{\ast}-t_{p}^{\dag}\hat{\rho}\hat{\rho}^{\mathrm{T}}t_{p}^{\mathrm{T}}\Big\}\;.
\end{eqnarray}
This can be rewritten as
\begin{equation}
\hat{T}_{\chi_{p}}=-\frac{k}{4}\left\{\left(-t_{p}^{\dag}\;r_{p}^{\dag}\right)\left(\begin{array}{cc}\hat{\rho}\hat{\rho}^{\mathrm{T}}&\hat{\rho}\hat{\theta}^{\mathrm{T}}\\
\hat{\theta}\hat{\rho}^{\mathrm{T}}&\hat{\theta}\hat{\theta}^{\mathrm{T}}\end{array}\right)\left(\begin{array}{c}t_{p}\\-r_{p}\end{array}\right)+
\left(-t_{p}^{\mathrm{T}}\;r_{p}^{\mathrm{T}}\right)\left(\begin{array}{cc}\hat{\rho}\hat{\rho}^{\mathrm{T}}&\hat{\rho}\hat{\theta}^{\mathrm{T}}\\
\hat{\theta}\hat{\rho}^{\mathrm{T}}&\hat{\theta}\hat{\theta}^{\mathrm{T}}\end{array}\right)\left(\begin{array}{c}t_{p}^{\ast}\\-r_{p}^{\ast}\end{array}\right)\right\}\;,
\end{equation}
which equals
\begin{eqnarray}
\hat{T}_{\chi_{p}}=\frac{k}{4}\left\{\mu_{p}^{\dag}G\hat{\scripty{r}}\;\hat{\scripty{r}}^{\mathrm{T}}G\mu_{p} +\mu_{p}^{\mathrm{T}}G\hat{\scripty{r}}\;\hat{\scripty{r}}^{\mathrm{T}}G\mu_{p}^{\ast}\right\}=\qquad\qquad\qquad\qquad\qquad\qquad\nonumber\\
\frac{k}{2}\left\{\mu_{p}^{\dag}G\hat{\scripty{r}}\;\mu_{p}^{\mathrm{T}}G\hat{\scripty{r}}+\mu_{p}^{\mathrm{T}}G\hat{\scripty{r}}\;\mu_{p}^{\dag}G\hat{\scripty{r}}\right\}=
\frac{1}{2}\left(\hat{a}_{p}^{\dag}\hat{a}_{p}+\hat{a}_{p}\hat{a}_{p}^{\dag}\right)\;,
\end{eqnarray}
where, we used that $\hat{\scripty{r}}^{\mathrm{T}}G\mu_{p}$ is scalar so that $\hat{\scripty{r}}^{\mathrm{T}}G\mu_{p}=\big(\hat{\scripty{r}}^{\mathrm{T}}G\mu_{p}\big)^{\mathrm{T}}=-\mu_{p}^{\mathrm{T}}G\hat{\scripty{r}}$.
\end{subappendices}

\end{document}